\documentclass[a4paper]{article}

\usepackage{INTERSPEECH2022}
\usepackage{xcolor}
\usepackage{float}
\usepackage{makecell}

\title{Non-Parallel Voice Conversion for ASR Augmentation
}
\name{Gary Wang, Andrew Rosenberg, Bhuvana Ramabhadran, Fadi Biadsy, Yinghui Huang, Jesse Emond, Pedro Moreno Mengibar}

\address{
  Google \textsuperscript{1}}
\email{\{wgary,rosenberg,biadsy,bhuv,huangyinghui,emond,pedro\}@google.com}

\begin{document}

\maketitle
\begin{abstract}

Automatic speech recognition (ASR) needs to be robust to speaker differences.  Voice Conversion (VC) modifies speaker characteristics of input speech. This is an attractive feature for ASR data augmentation.  In this paper, we demonstrate that voice conversion can be used as a data augmentation technique to improve ASR performance, even on LibriSpeech, which contains 2,456 speakers. For ASR augmentation, it is necessary that the VC model be robust to a wide range of input speech.  This motivates the use of a non-autoregressive, non-parallel VC model, and the use of a pretrained ASR encoder within the VC model. This work suggests that despite including many speakers, speaker diversity may remain a limitation to ASR quality.   Finally, interrogation of our VC performance has provided useful metrics for objective evaluation of VC quality.

\end{abstract}

\section{Introduction}

It is critical for automatic speech recognition (ASR)  to be robust to speaker differences.  This is typically addressed by training on speech from a wide variety of speakers.  In this paper, we show that despite being trained on ~1000 hours of speech, augmenting speech with diverse speaker characteristics can improve speech recognition performance. 

Voice conversion (VC) models convert input speech from its source speech to some target speaker, transferring the speaker timbre and other characteristics while retaining the lexical content of the source speech.  While there has been successful approaches to voice conversion in recent years (Section \ref{sec:related-work}), VC performance is typically evaluated on clean, in-domain speech.  That is, the same style of speech used in training the VC model is used for evaluation. We have found that this can lead to VC systems that fail to generalize to speech and speakers that are substantially different than the training material  (Section \ref{ss:vc-conversion-quality}).  

Since ASR training data is typically more diverse than VC training data, constructing a robust VC system is essential for ASR augmentation.  ASR systems have to interact with speech from a variety of recording conditions and a broad range of speakers including speakers with less common accents.

In service of this robustness we make three important VC design choices. First, we pursue a non-parallel voice conversion approach (Section \ref{sec:non-parallel-vc}).  By not requiring parallel data for training, we can train on much more data than we could otherwise. Second, we use a non-autoregressive decoder. We found that autoregressive models were susceptible to attention failures which were catastrophic for their application as a data augmentation technique. Third, we initialize the VC encoder with a pre-trained ASR encoder.  This choice substantially improves the robustness of the VC model to diverse inputs (Section \ref{sec:experiments}), while moderately limiting VC naturalness.  For ASR augmentation, we find that this substantial improvement to robustness is more important than a marginal improvement to naturalness.

The main contribution of this paper are as follows:
\begin{itemize}
    \item We show that voice conversion can be used to successfully augment speech recognition training data. We find relative word error rate (WER) improvements up to ~$6\%$ when augmenting LibriSpeech which contains 2,456 speakers.
    \item We show that we can substantially improve the robustness of voice conversion systems to unseen corpora and different speaker characteristics by initializing the VC encoder with an ASR encoder.
    \item We demonstrate VC metrics that can be used for model selection prior to subjective listening tests.
\end{itemize}

\begin{center}
\begin{figure*}[t]
  \centering
  \includegraphics[width=\linewidth]{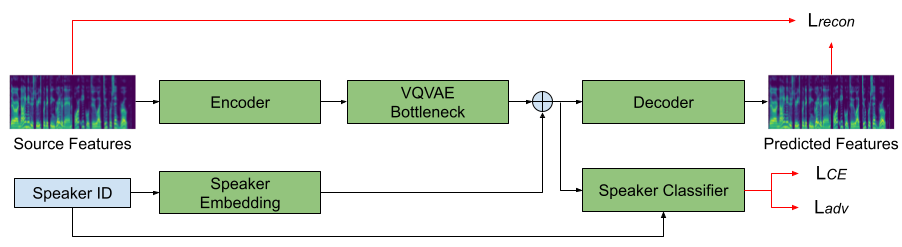}
  \caption{Schematic diagram of proposed voice conversion system.}
  \label{fig:vs-model}
\end{figure*}
\vspace{-3mm}
\end{center}

\section{Related Work}
\label{sec:related-work}
Non-parallel voice conversion can be broadly categorized into two major approaches, the first using Phone posteriogram (PPG) features \cite{sun2016phonetic,zhang2019non}. PPGs acts as speaker-invariant representations that can be easily utilized to achieve conversion. The other main approach is with auto-encoding style training approaches, utilizing both variational auto-encoder (VAE)\cite{hsu2016voice,huang2020unsupervised,tobing2019non,kameoka2019acvae,qian2019autovc,qian2020unsupervised} and vector-quantized VAE (VQ-VAE) \cite{ding2019group,van2020vector,zhang2020neteasegames,kobayashi2021crank} approaches. The auto-encoding style training approaches try to disentangle speaker speaker information from the content of source speech, using various methods from bottlenecking to adversarial training to prevent speaker leakage in the content or speech encoder. In \cite{qian2019autovc}, instead of adversarial training, carefully designed bottlenecks were coupled with auto-encoding loss to train the VC system. In \cite{kobayashi2021crank}, this is achieved by VQ-VAE of the speech encoder outputs.

Recent approaches have utilized pre-trained speech encoder to extract speaker agnostic representation. In \cite{zhao2018accent}, ASR trained to predict PPGs are used on source speech to achieve accent conversion. In \cite{wang2021enriching}, a trained ASR encoder is used as the speech encoder, along with explicit prosody modelling to achieve speaker and prosody in VC. In \cite{wang2021accent}, accent and speaker disentanglement and control in VC was achieved in stages, where first stage is accent-agnostic ASR training, from which the trained speech encoder is used to further conduct VC training and disentangle accent and speaker.

Previous work on voice conversion and morphing augmentations to help augment ASR training include Vocal Tract Length Perturbation (VTLP) \cite{cui2015data,bellegarda1994metamorphic}. More recent approaches use neural VC models to augment speech. In \cite{shahnawazuddin2020voice}, a VC model is used to augment adult speech into children speech to help with children speech ASR. However in this work, the VC model was a Cycle-GAN\cite{zhu2017unpaired} architecture trained on the same in-domain data on which the ASR training was also conducted with. In \cite{baas2021voice}, a VC system was utilized as augmentation for low resource languages. This system utilizes representations from a pretrained encoder with CPC\cite{nguyen2020zero} objective, instead of one trained on ASR targets. The VC model augments speech in the very low data regime.

\vspace{-2mm}
\section{Non-Parallel Voice Conversion}
\vspace{-2mm}
\label{sec:vc4asr}
\label{sec:non-parallel-vc}
For our proposed VC model, we utilize an architecture consisting of an encoder, a bottleneck layer, and a non auto-regressive decoder. 
%
\label{s:vc-training}
Figure \ref{fig:vs-model} shows the architecture of the VC model.

\textbf{Encoder:} The VC encoder is similar to the encoder in the Conformer\cite{gulati2020conformer} ASR model. Specifically, the encoder consists of sub-sampling convolution layer that reduces the time dimension of the input speech by 4x. This is then followed by a stack of conformer layers (16 layers), followed by a output layer norm layer. We study two versions of encoders, the first being an conformer encoder that is trained from scratch jointly with the rest of the system. The second version being a frozen encoder, where the weights are obtained from an ASR model trained on 960 hours of Librispeech.

\textbf{Speaker Embedding:} Speaker information is provided in the form of one-hot vectors, which are passed through a learned speaker embedding that maps from one-hot vectors to a dense 256-dim embedding vector.

\textbf{Decoder:} The decoder is a non-autoregressive decoder composed of RNN layers followed by convolution layers. Speaker embedding is provided along with the encoder output to the decoder. Specifically, the dense speaker embedding vector is copied $T$ times across time and then concatenated together with encoder outputs to be fed to the decoder. The decoder consists of 2 bi-directional LSTM layers of size 1024 followed by a stack of convolution layers.  During development, we also explored an auto-regressive, attention-based decoder.  However, we found that during inference even within-domain speech would sometimes show attention failures leading to extremely poor conversion \cite{non-ar-tacotron}.  While there are ways to mitigate these failures, since robustness is necessary for ASR augmentation we pursued the less error-prone, non-autoregressive approach.

\textbf{VQVAE Bottleneck:}
A Vector Quantized Variational Auto Encoding (VQVAE) Bottleneck \cite{van2017neural} is used to quantize encoder outputs into discrete codebook entries. This restricts information through the VC system during auto-encoding training. 
The VQVAE bottleneck typically contain 3 loss terms, the reconstruction loss, codebook loss and commitment loss.

\begin{eqnarray}
    L_{codebook}&=&||sg[z_e(x)] - e||^{2}_{2}\\
    L_{commit}&=&\beta||z_e(x) - sg[e]||_{2}^{2}
\end{eqnarray}

Reconstruction loss is computed on the decoder, thus we simply add codebook and commitment loss from VQVAE to the overall loss objective during training. We fix VQVAE codebook size to 128 and group size to be 2.

\textbf{Adversarial Speaker Classification:}
To disentangle speaker information, we apply adversarial speaker classification loss on the VQVAE codebook outputs to further ensure no speaker leakage. Given encoder outputs $m_{enc}(x)$ and speaker labels $l_{speaker}$, we combine reverse gradient operation $R_{grad}$ along with softmax cross entropy loss as follows:
\begin{equation}
    L_{adv} = CE(R_{grad}(m_{enc}(x)), l_{speaker})
\end{equation}

In effect, the training will try to maximize $L_{adv}$, which in effect will remove as much speaker information from the VQVAE output as possible. We scale adversarial gradient by weight, and find that this adversarial gradient weight term requires careful tuning, too small and it does nothing, too large and it overwhelms the reconstruction loss, see section \ref{ss:vc-quaity-metrics} for details.

\textbf{Objective Function}
\label{vc-training-setup}
We train on non-parallel data, and only utilize speech features and speaker embedding for the loss computation. To enable the VC models to be used as ASR data augmentation, we train our VC model to predict ASR log-mel features directly. During VC training, the same log-mel features are used as input and target. We use Huber loss for reconstruction loss. The model is jointly optimized by minimizing $L_{total}$, with all loss weights set to 1.0.

\begin{eqnarray}
L_{recon}&=&
    \left\{\begin{matrix}
        \frac{1}{2}(y - \hat{y})^{2} & if \left | (y - \hat{y})  \right | < \delta\\
        \delta ((y - \hat{y}) - \frac1 2 \delta) & otherwise
    \end{matrix}\right. \\
    L_{total}&=& L_{recon} + \gamma L_{codebook} + \epsilon L_{commit} + \eta L_{adv}
\end{eqnarray}

\textbf{Frozen ASR Encoder}
\label{ss:frozen-asr}
In addition to training the full VC model (Figure \ref{fig:vs-model}) from scratch, we replace the encoder network with the encoder of an ASR model.  ASR models implicitly normalize out speaker differences. ASR outputs are not (explicitly) dependent on speaker characterstics so they are less likely to be retained in the internal representations of the ASR network.  We use an ASR encoder as the VC encoder, and freeze its weights, training only the decoder, VQVAE and speaker embedding.
By freezing the encoder, it retains the representations learned during ASR training which leads to improved robustness.

\section{Data}
\textbf{VC Corpora:}
\label{ss:vcdata}
Three TTS corpora were used for training the VC model. The first {\bf LS} is the publicly available LibriSpeech Corpus containing 960 Hours of speech with 2,456 speakers~\cite{panayotov2015librispeech}. 
The second {\bf EN-VC} is an in-house corpus containing professional English speakers from 6 English locales, including United States (en/us), Britain (en/gb), India (en/in), Singapore (en/sg), Nigeria (en/ng) and Australia (en/au). This corpus contains 58 speakers, totaling around 400 hours of studio quality speech intended for Text-to-speech (TTS) training and research and hence was recorded at 48KHz. The third corpus {\bf SV-VC} is, similar to the second, an in-house Swedish (sv/se) dataset collected for TTS training and research, which contains 21 hours of across 6 speakers. The data distribution between speakers is skewed, with speaker hours ranging from 0.5 hours to 40 hours. All datasets are downsampled to 16kHz, and we use 80-dim mel spectrum as features, with a frame size of 25ms and  a 10ms  frame shift.

\textbf{ASR Corpora:}
\label{ss:asrdata}
For ASR, three training datasets were used in this work. The first is {\bf LS} the public  LibriSpeech\cite{panayotov2015librispeech} corpus. The other two datasets are in-house data sets comprising of anonymized short utterances representative of Google's voice search (VS) traffic in two languages, English {\bf EN-ASR} and Swedish {\bf SV-ASR}. While the choice of English allows us to analyze two very different domains, namely, audio books and voice search with similar amounts of training data(~1000 hours), the choice of Swedish(~6500 hours) allows us to explore some transfer effects. While the {\bf LS} corpus used in VC and ASR training is identical, the in-house VC ({\bf EN-VC} and {\bf SV-VC}) and ASR corpora ({\bf EN-ASR} and {\bf SV-ASR}) are completely disjoint.

\section{Experiments \& Results}
\label{sec:experiments}
\subsection{VC Model Training}
\label{sss:vctrain}
The VC model as described in Section \ref{s:vc-training} is trained with constant learning rate of 1e-4 for 800k steps with Adam optimizer. Note that the VC model is trained on studio quality professional speech and has not seen any ASR training data. When using a frozen ASR encoder, the encoder is initialized from a LibriSpeech trained ASR encoder.
For the LibriSpeech ASR experiments, the VC model is trained on Librispeech dataset. For in-house datasets \textbf{EN-ASR} and \textbf{SV-ASR}, we train VC models on our respective internal TTS datasets (Section \ref{ss:vcdata}).

For \textbf{LS-ASR} experiments, the VC model is trained on Librispeech data. For \textbf{EN-ASR} experiments, the VC model is trained on our internal en-us TTS data. For \textbf{SV-ASR} experiments, we train VC model on both our internal US and Swedish TTS datasets to have access to more speakers.  Note, that the SV VC model is initialized with the LibriSpeech ASR encoder.  This investigates the use of an ASR encoder as a proxy for a language agnostic representation for VC.

\subsection{VC for ASR Augmentation}
\label{sss:vc_aug}

When using VC as an ASR augmentation technique, we follow \cite{wang2020improving,chen2020improving} for ASR training.  Two views of the same data is presented to the ASR model, original and VC converted speech. Additionally a decoder consistency term is introduced to force model prediction to be consistent for the two views.

We apply SpecAugment to both views.   During VC inference, we uniform randomly sample speaker ids from the pool of {\bf VC training} speakers whether {\bf LS}, {\bf EN-VC} or {\bf SV-VC} to generate augmented versions of the ASR training data. No speaker information from the ASR training data is used, nor is it available for {\bf EN-ASR} or {\bf SV-ASR}. The VC model is frozen during ASR training and used purely as an augmentation technique.

\subsection{VC Quality Analysis}
\label{ss:vc-conversion-quality}

For analyzing VC quality, we train two models on {\bf EN-VC}, one using a frozen ASR encoder (\textbf{ASR Encoder}) and one from scratch (\textbf{VC Encoder}).

We train a separate WaveRNN neural vocoder for each VC model that maps from predicted log-mel features to 16kHz waveform. We analyze conversion quality with 3 measures, Mean Opinion Score Naturalness (MOS), Speaker Similarity, and Word Error Rate (WER) scored with in-house ASR system as a proxy for intelligibility. MOS and Similarity are rated on a 5 point Likert scale. For all evaluations, we convert from 5 example utterances from 20 source speakers to 5 target speakers, totaling 500 utterances.

\begin{center}
\vspace{-3mm}
\label{tb:in-locale-mos}
\begin{table}[th!]
\caption{MOS naturalness, Speaker Similarity and conversion word-error-rate for In-Domain, In-Locale Conversion}
\begin{tabular}{l|c|c|c} \toprule
    {Method} & {MOS} & {Similarity} &{WER} \\ \midrule\midrule
    \begin{tabular}[l]{@{}l@{}}Groundtruth \end{tabular}  & 4.347 ± 0.049  & - & 10.9 \\\midrule
    \begin{tabular}[l]{@{}l@{}}\makecell{VC Encoder} \end{tabular}   & 4.001 ± 0.069  & 4.227 ± 0.059 & 14.4 \\\midrule
    \begin{tabular}[l]{@{}l@{}} \makecell{ASR Encoder} \end{tabular}   & 3.611 ± 0.075 & 4.419 ± 0.051 & 16.0 \\\midrule
\end{tabular}
\end{table}
\vspace{-4mm}
\end{center}

\textbf{In-Domain Conversion}
\label{sss:in-domain-conversion}
Our first analysis evaluates in-domain conversion, where source speakers and target speakers come from the same US English locale and corpus ({\bf EN-VC}) (Table  \ref{tb:in-locale-mos}). When comparing VC using a trained encoder to the Frozen ASR encoder, we find a higher MOS and lower WER from the converted speech, but we find lower speaker similarity.  This confirms the hypothesis that the frozen ASR encoder reliably eliminates speaker information from its output representation.

\textbf{Cross-Locale Conversion}
\label{sss:cross-domain-conversion}
To evaluate cross-local conversion, the source speakers consists of 5 non-US English locales (see Section \ref{ss:vcdata}), converting to 5 US English speakers.  Since ASR training and evaluation data often includes a variety of accents, this evaluates how robust our VC system will be to differently accented speech.

Table \ref{tb:cross-locale-mos} shows that both models result in less natural conversion (MOS) though the ASR Encoder mitigates this regression.  Moreover, the ASR Encoder shows better robust speaker conversion (Similarity) and better WER.

\begin{center}
\vspace{-4mm}
\begin{table}[th!]
\caption{ \label{tb:cross-locale-mos}
MOS naturalness, Speaker Similarity and conversion word-error-rate for Cross-Locale Conversion}
\begin{tabular}{l|c|c|c} \toprule
    {Method} & {MOS} & {Similarity} &{WER} \\ \midrule\midrule
    \begin{tabular}[l]{@{}l@{}}Groundtruth \end{tabular}  & 4.347 ± 0.049  & - & 12.0 \\\midrule
    \begin{tabular}[l]{@{}l@{}}\makecell{VC Encoder} \end{tabular}   & 3.344 ± 0.094  & 2.629 ± 0.086 & 39.4  \\\midrule
    \begin{tabular}[l]{@{}l@{}} \makecell{ASR Encoder} \end{tabular}   & 3.281 ± 0.088 & 3.494 ± 0.098 & 34.7\\\midrule
\end{tabular}
\vspace{-4mm}
\end{table}
\vspace{-2mm}
\end{center}

\textbf{Out-Of-Corpus Conversion}
\label{sss:ood-conversion}
For out-of-corpus conversion, the source speech consists of 20 random speakers from LibriTTS test-other data. The VC training data is {\bf EN-VC}. 
Table \ref{tb:ood-mos} shows that when encountering out-of-domain input speech, the VC Encoder system attains worse MOS score and much higher conversion WER as compared to the ASR Encoder system.  We note that while the VC system was trained on {\bf EN-VC}, the ASR Encoder has only seen LibriSpeech training utterances.

\begin{center}
\vspace{-4mm}
\begin{table}[th!]
\caption{ \label{tb:ood-mos}MOS naturalness, Speaker Similarity and conversion word-error-rate for Out-of-Corpus Conversion}
\begin{tabular}{l|c|c|c} \toprule
    {Method} & {MOS} & {Similarity} &{WER} \\ \midrule\midrule
    \begin{tabular}[l]{@{}l@{}}Groundtruth \end{tabular}  & 4.347 ± 0.049 & - & 10.4 \\\midrule
    \begin{tabular}[l]{@{}l@{}}\makecell{VC Encoder} \end{tabular}   & 2.336 ± 0.087  & 2.549 ± 0.08 & 49.2  \\\midrule
    \begin{tabular}[l]{@{}l@{}} \makecell{ASR Encoder} \end{tabular}   & 2.982 ± 0.088 & 4.068 ± 0.071 & 21.0  \\\midrule
\end{tabular}
\vspace{-4mm}
\end{table}
\vspace{-4mm}
\end{center}
These three experiments demonstrate that the use of a pretrained, frozen ASR encoder in VC leads to substantial improvements to VC robustness to accent and recording condition.

\subsection{VC Augmentation \& ASR Training}
We now present our results of using VC as general ASR augmentation strategy across 3 datasets (Section \ref{ss:asrdata}): {\bf LS}, {\bf EN-ASR} and {\bf SV-ASR}.  
Due to the low quality conversion of out-of-locale, and out-of-corpus sampled, all VC systems across 3 experiments use a Librispeech-trained {\bf ASR encoder} as the VC encoder. 

\label{sss:librispeech}
For \textbf{LibriSpeech} experiments, we use a conformer RNN transducer model as the ASR model. The ASR model consists of 16 layers of conformer layers in the encoder, each with conformer dimension of 16 and 4 attention heads. The decoder consists of a RNN transducer with RNN dimension of size 320. The system is trained for up to 120k steps with Adam optimizer, with annealed learning rate to 5e-4. VC augmentation is described in section \ref{sss:vc_aug}.
Table \ref{tb:ls} shows that VC augmentation improves WER by 0.2 and 0.1 absolute on test-clean and test-other. The LibriSpeech training data contains speech from 2,456 speakers, and still the augmentation provided by VC within a consistent data augmentation framework \cite{wang2020improving} is able to provide improved performance.

\begin{table}[th!]
\caption{ \label{tb:ls}VC Augmentation Results on Librispeech}
\begin{tabular}{c|c|c} \toprule
    {Method} & { LS test} & {LS test-other}  \\ \midrule\midrule
    \begin{tabular}[l]{@{}l@{}} Baseline \end{tabular} & 3.0  & 6.8   \\\midrule
    \begin{tabular}[l]{@{}l@{}} + VC Augmentation \end{tabular} & 2.8 & 6.7   \\\midrule
\end{tabular}
\centering
\end{table}

For \textbf{EN-ASR}  experiments, the ASR system consists of a Conformer \cite{gulati2020conformer} RNN transducer model of approximately 600 Million parameters. The ASR encoder consists of 24 layers of conformer layers each with 1024 dimension and 8 attention heads. The decoder consists of a RNN transducer with 2 Bi-directional LSTM layers each with 2048 dimension. The encoder has been pre-trained via self-supervised learning on LibriLight data along with unspoken text injection via tts as laid out in \cite{chen2021injecting}. The baseline system is fine-tuned on {\bf EN-ASR}. 

The ASR model used in the Swedish ASR (\textbf{SV-ASR}) experiment is a Conformer transducer model of approximately 120 Million parameters. The encoder consists of 12 Conformer\cite{gulati2020conformer} layers, each with conformer dimension of 512 encoder dimension and 8 attention heads. The decoder consists of an embedding based decoder \cite{ghodsi2020rnnt}, with embedding dimension of 640 dimension. The model is trained with random initialization with Adam optimizer for 300k steps, with a learning rate of 5e-4. VC augmentation is utilized in the set up as described in section \ref{sss:vc_aug}.  The VC model is trained on both {\bf EN-VC} and {\bf SV-VC} data.  When initializing the VC encoder with an ASR Encoder, this is the same Librispeech-trained ASR Encoder used in the {\bf EN-ASR} models.  This addresses the question of whether an out-of-language ASR model can be used within a VC model.  This is akin to performing PPG Voice Conversion with a phone recognition model trained on a different language.

Table \ref{tb:envs} shows VC augmentation to be an effective augmentation technique for {\bf EN-ASR} fine-tuning, improving WER from 6.7\% to 6.3\%, while the improvement is smaller for {\bf SV-ASR} at only 1\%.

\begin{table}[th!]
\vspace{-2mm}
\caption{\label{tb:envs} \label{tb:svvs} VC Augmentation Results on {\bf EN-ASR} and {\bf SV-ASR}}
\begin{tabular}{c|c|c} \toprule
    {Method} & { EN-ASR WER} & {SV-ASR WER} \\ \midrule\midrule
    \begin{tabular}[l]{@{}l@{}} Baseline \end{tabular} & 6.7 & 13.7  \\\midrule
    \begin{tabular}[l]{@{}l@{}} + VC Augmentation \end{tabular} & 6.3 & 13.6  \\\midrule
\end{tabular}
\centering
\vspace{-3mm}
\end{table}

\subsection{VC Training Quality Metrics}
\label{ss:vc-quaity-metrics}
Tuning weights for adversarial losses can be tricky, too small, they do nothing, too large, the main loss (here reconstruction) is overwhelmed.
During our VC modelling development, we found two measures to be useful for tracking VC training performance. Table \ref{tb:adv-and-perplexity} shows VC training runs with different adversarial gradient weights. Also provided is the speaker classification accuracy, VQVAE codebook perplexity and the reconstruction loss.  Based on these three measures, we identify good candidate VC models as those with low speaker accuracy, high VQVAE perplexity and low reconstruction loss.  While not narrowly predictive of VC performance, these three measures were able to identify models that were performing VC well enough to be evaluated by subjective tests.

\begin{center}
\label{tb:adv-and-perplexity}
\begin{table}[th!]
\caption{The effect of adversarial weight on metrics available during VC training.}
\begin{tabular}{c|c|c|c} \toprule
    {\makecell{Speaker Adversarial \\ Gradient Weight}} & {\makecell{Speaker \\ Acc}} & {\makecell{VQVAE \\ Perplexity}} & {\makecell{Reconstruction \\ Loss}} \\ \midrule\midrule
    \begin{tabular}[l]{@{}l@{}}\end{tabular}0.0 & 80\% & 110 & 0.038  \\\midrule
    \begin{tabular}[l]{@{}l@{}}\end{tabular} \textbf{0.1} & \textbf{12\%} & \textbf{105} & \textbf{0.045} \\\midrule
    \begin{tabular}[l]{@{}l@{}}\end{tabular} 0.5 & 11\% & 80 & 0.070  \\\midrule
    \begin{tabular}[l]{@{}l@{}}\end{tabular} 1.0 & 10\% & 60 & 0.085 \\\midrule
    
\end{tabular}
\end{table}
\vspace{-4mm}
\end{center}

\section{Conclusion}
We demonstrate that even when trained on more than two thousand speakers, LibriSpeech ASR performance can be improved by Voice Conversion augmentation. We find similar ($\sim6\%$) relative gains on an in-house English ASR task, but smaller improvements ($\sim1\%$) on a comparable Swedish task.  It is necessary for VC models to be robust to noisy inputs and diverse speakers to achieve these result. We show that using a pre-trained ASR Encoder to serve as the encoder in a non auto-regressive, non-parallel VC model is a successful approach to promoting the necessary VC robustness.

\bibliographystyle{IEEEtran}

\bibliography{main}

\end{document}